\documentclass[letterpaper]{jpconf}
\usepackage{graphicx}

\begin{document}

\title{Long-range atom-surface interactions for cold atoms}

\author{James F. Babb}

\address{Institute for Theoretical Atomic, Molecular,
and Optical Physics, Harvard-Smithsonian Center for Astrophysics,
MS 14, 60 Garden St., Cambridge, MA 02138 USA}

\ead{jbabb@cfa.harvard.edu}

\begin{abstract}
  Studies of the long-range interactions between cold atoms and
  surfaces are now of vital interest. The interest is partly driven by
  nanotechnology applications, partly by the exploding interest in the
  encompassing superfield of Casimir effects, and partly by the
  burgeoning overlap between atomic and molecular physics, condensed
  matter, and quantum optics.  This tutorial lecture will address
  long-range atom-surface interactions for cold atoms, including an
  overview of Casimir-Polder interactions and their various
  manifestations.  Some previous theoretical studies that are of
  particular relevance will be reviewed.  In addition some different
  approaches to the problem and corresponding results, especially
  concerning the effects of substrate composition, geometry, and
  finite temperature, will be discussed.
\end{abstract}

\section{Introduction}

The interactions between atoms and surfaces are important in many
areas of physics. Consequently, the literature is extensive and broad.
In this tutorial, I will focus primarily on the long-range
atom-surface interactions relevant to the atomic, molecular, and
optical physics of cold atoms near surfaces.  I will introduce the
short-range Lennard-Jones potential and the long-range retarded
Casimir-Polder potential, discuss the accurate calculation of the
interaction coefficients and the calculation of the potentials for
distances that are not too close to the surfaces, and the inclusion of
real surface properties, such as dielectric response and temperature.
The paper will cover theoretical aspects. A number of recent
experiments were presented at the \textit{Conference on Atoms and
  Molecules near Surfaces}, see the papers in the present volume, and
Refs.~\cite{AspDal02,BloDuc05}.

The interaction potential between a ground state atom and a perfectly
conducting wall is
\begin{equation}
\label{small-R}
V (R)  =    -C_3 R^{-3} ,
\end{equation}
where $R$ is the distance between the atom and the wall, $C_3$ is the
Lennard-Jones coefficient, and atomic units are used throughout
(except when it is useful to exhibit $\hbar$ and the speed of light
$c$).  The interaction between polarizable systems is mediated by the
exchange of virtual photons\footnote{The \textit{resonant} interaction
  between a ground state atom and an excited state atom, \textit{e.g.}
  a Na(3s) atom and a Na(3p) atom, will not be considered here. }  and
as the separation of the systems increases there is a characteristic
change in the interaction energy between them.  In 1948, Casimir and
Polder showed with quantum electrodynamics that for an atom
interacting with a perfectly conducting wall the potential
is~\cite{CasPol48}
\begin{equation}
\label{CP}
V(R) = - \textstyle{\frac{3}{8\pi}}\hbar c \alpha_d(0)R^{-4}, 
\qquad R\rightarrow\infty , 
\end{equation}
where $\alpha_d (0)$ is the static electric dipole polarizability of
the atom. Subsequent to Casimir and Polder, the result (\ref{CP}) has
been obtained by many other investigators, see, for example, the
comprehensive treatments in 
Refs.~\cite{Boy73,MarToi82a,WylSip84,Hin92,ZhoSpr95,BuhKnoWel04,AntPitStr04}.

The appearance of $\hbar$ and $c$ and the weaker interaction (higher
inverse power of $R$) are signatures of the Casimir-Polder potential.
To understand the interaction potential~(\ref{CP}) it is convenient to
begin with a simple derivation of the retarded potential between two
finite, polarizable systems that gives the correct expressions up to
numerical constants. I follow the simple physical approaches presented
in Refs.~\cite{SprKel78,Mil93,Spr96}; other approaches 
can be found in Refs.~\cite{Boy69} and \cite{PowThi93b}.

Suppose that the two polarizable systems (either might be one of atom,
electron, ion, or wall) are in the presence of a uniform background
(vacuum) field $\textbf{E}_b (\omega)$.  The interaction energy of the
two systems is given by
\begin{equation}
\label{energy}
U(\omega,R) \sim \alpha_1(\omega)
     [\textbf{E}_{2 \rightarrow  1}  (\omega,R) +
     \textbf{E}_b (\omega) ]^2 + {1 \leftrightarrow  2},
\end{equation}
where the second term signifies the interchange of system 1 and system
2 in the first term and at large distances the electric field at
system 1 coming from system 2 is
\begin{equation}
\textbf{E}_{2 \rightarrow  1} (\omega,R) \sim
     e^{iR\omega/c} \frac{\omega^{2}}{c^{2}R} \textbf{p}_2(\omega),
\end{equation}
where the  electric dipole moment is 
\begin{equation}
\textbf{p}_2 (\omega) =\alpha_2 (\omega) \textbf{E}_b (\omega) 
\end{equation}
and the dynamic electric dipole polarizability is
\begin{equation}
\label{polariz}
\alpha (\omega) = \sum_u f_{u}/[(E_u-E_0)^2-\omega^2],
\end{equation}
with $f_u$  the oscillator strength of state $u$ and $E_u-E_0$ 
the transition frequency between the states $u$ and $0$.
Note that the ``sum'' in eq.~(\ref{polariz}) includes
a sum over all discrete transitions and an
integration over the continuum. 

To obtain the $R$-dependent interaction energy,  we retain only 
the cross terms in eq.~(\ref{energy}) giving
\begin{equation}
U(\omega,R) \sim \alpha_1 (\omega)   {\bf
 E}_{2 \rightarrow 1}(\omega,R) \textbf{E}_b (\omega) 
\sim
\alpha_1(\omega)  \alpha_2(\omega)\textbf{E}_b^2 (\omega) 
e^{iR\omega/c} (\omega^2/c^2 R) .
\end{equation} 

Summing over modes of the background field, replacing
$\sum_\textbf{k}$ with ${\rm V}\int d\omega \omega^2/c^3$ and
$\textbf{E}_b^2(\omega)$ with $\hbar \omega/{\rm V}$, 
and cutting off the integration at the
highest relevant characteristic frequency, 
a simple integral expression for
the potential is 
obtained~\cite{SprKel78} 
\begin{equation}
\label{physical}
U(R) \sim \frac{\hbar}{c^{5}R} \int_{0}^{c/R} d\omega
   \omega^{5} \alpha_1(\omega)\alpha_2 (\omega) .
\end{equation}

This useful formula, eq.~(\ref{physical}), can reproduce the asymptotic
Casimir-Polder potentials for various cases such as the interaction
between two atoms or between and atom and a wall; its range of
validity can be extended~\cite{SprBabZho94,Spr96}.  In treating the
\textit{asymptotic} potential using eq.~(\ref{physical}), we  make the
replacement $\alpha(\omega)\rightarrow \alpha(0)$.  For example, the
retarded Casimir-Polder potential between two atoms is, according to
eq.~(\ref{physical}), $U (R) \sim \hbar c \alpha_1 (0) \alpha_2 (0)
R^{-7}$ in agreement with Casimir and Polder~\cite{CasPol48} who
obtained
\begin{equation}
\label{CP-atom-atom}
U(R) = -\textstyle{\frac{23}{4\pi}}
\hbar c \alpha_1 (0) \alpha_2 (0) R^{-7} ,
\quad R\rightarrow \infty, 
\end{equation}

Spruch and Kelsey~\cite{SprKel78} showed how eq.~(\ref{physical}) can
reproduce the atom-wall interaction. Let the wall be approximated by a
sphere of radius $CR$, where $C$ is a number less than one, perhaps
around $\frac{1}{4}$, so that the systems are separated by a total
distance $R+CR=R(1+C)$. Then the polarizability of the sphere (wall)
is $(CR)^3$, which when substituted into eq.~(\ref{physical}) with the
replacement of $R$ by $R(1+C)$ (reasonable 
since $C \ll 1$) yields,
\begin{equation}
\label{simple-wall}
V (R) \approx C^3 \alpha_d (0) \hbar c R^{-4},
\end{equation}
in agreement with eq.~(\ref{CP}).

Can an expression more precise than eq.~(\ref{simple-wall}) for the
coefficient in the asymptotic atom-wall interaction potential be
obtained simply by integrating the asymptotic atom-atom potential
eq.~(\ref{CP-atom-atom}) over all the atoms constituting the wall? The
calculation was carried out in Ref.~\cite{MilShi92} with the result
\begin{equation}
V(R) \approx -\textstyle{\frac{69}{160\pi}}  \hbar c \alpha (0) R^{-4},
\end{equation}
which is about 15\% larger than the result of Casimir and Polder,
eq.~(\ref{CP}).  [A similar discrepancy arises when the coefficient
$C_3$ appearing in eq.~(\ref{small-R}) is estimated by integrating the
$R^{-6}$ van der Waals interaction pairwise between an atom and each
of the atoms in the wall~\cite{FowHut86}.]  The overestimation of the
actual interaction coefficient is attributed to the non-additivity of
long-range dispersion forces---the pairwise treatment does not account
for three-body and higher-order interactions. The treatment of 
walls will considered in sections~\ref{ideal} and
\ref{beyond-ideal} below.

\section{Coefficients}

We now leave the Casimir-Polder potential until section~\ref{ideal}
and consider the accurate evaluation of the Lennard-Jones atom-wall
interaction potential~eq.(\ref{small-R}) for separations sufficiently
large that the exchange energy of the overlap between the atomic and
surface wave functions (cf.~\cite{VidIhmKim91}) is not important.  For
a perfectly conducting wall, the coefficient $C_3$ 
can be written as
\begin{equation}
\label{C3}
C_3 = \frac{1}{4\pi} \int_0^\infty \,d\omega \alpha_d (i\omega) ,
\end{equation}
or, from direct integration of eq.~(\ref{C3}), as 
\begin{equation}
\label{C3-direct}
C_3 = \frac{1}{12} 
   \left \langle 0 \left| \left(\sum_{i=1}^{N_e} \textbf{r}_i \right)^2
          \right| 0 \right\rangle ,
\end{equation}
where $|0\rangle$ is the wave function of the atom, $N_e$ is the
number of electrons, and $\textbf{r}_i$ is the position vector from the
nucleus to electron $i$.

The $C_3$ coefficients can be calculated in many ways, including
\textit{ab initio} methods~\cite{YanBab98,YanDalBab97}, density
functional theory (DFT)~\cite{HulRydLun99}, and semiempirical
methods~\cite{ColBar86,FowHut86,VidIhmKim91,KhaBabDal97}.  For H, the value of
$C_3$ is $\frac{1}{4}$~\cite{MarDalBab97}.  Accurate values of $C_3$
have been obtained using \textit{ab initio} non-relativistic methods
for Li~\cite{YanDalBab97} and He$(2\,^3\!S)$~\cite{YanBab98} and for
the heavy alkali-metal atoms (Na, K, Rb, Cs, and Fr) using
relativistic many-body perturbation theoretic
methods~\cite{DerJohSaf99,JohDzuSaf04}.
The most accurate $C_3$ values for some systems
of interest for cold atom studies are summarized in Table~\ref{C3-table}.
\begin{table}
\caption{\label{C3-table}Accurate values of the
coefficient $C_3$ for the interaction
of an atom in its
ground state (except He$(2\,^3\!S)$) 
with a perfectly conducting wall, eqs.~(\protect\ref{small-R})
and (\protect\ref{C3}), in atomic units.}
\begin{center}
\begin{tabular}{lllllllll}
\br
Atom & H   & He$(2\,^3\!S)$&Li    &Na  &K   & Rb & Cs & Fr \\
\mr
$C_3$&$\textstyle{\frac{1}{4}}$
            & 1.901         &1.518 &1.89&2.97&3.53&4.5 &4.71 \\
Ref. & \protect\cite{MarDalBab97}    
           & \protect\cite{YanBab98}
                        & \protect\cite{YanDalBab97} 
                            &\protect\cite{KhaBabDal97}
                            &\protect\cite{JohDzuSaf04}
                            &\protect\cite{JohDzuSaf04}
                            &\protect\cite{JohDzuSaf04}
                            &\protect\cite{JohDzuSaf04} \\ 
\br
\end{tabular}
\end{center}
\end{table}

Estimates of $C_3$ with eq.~(\ref{C3}) using a single oscillator
strength in eq.~(\ref{polariz}) can be inaccurate. For Li the value of
$C_3$ is about 1.518~\cite{YanDalBab97}, but the value obtained using
just the resonance transition is only 1.45, an underestimate of 4.5\%.
For Cs, the value of $C_3$ is about 4.45~\cite{JohDzuSaf04}, but the
value obtained using just the resonance transition is 2.59, an
underestimate of 42\%.  The discrepancy can be traced back to the
contribution of the atomic core electrons, see
eq.~(\ref{C3-direct})~\cite{DerJohSaf99}.  
Another way to see the origin of the discrepancy
is to consider eq.~(\ref{C3}), which is
an integral from $0<\omega<\infty$.
Recall that for high frequencies, 
\begin{equation}
\alpha_d (i\omega) \sim \sum_u f_{u} /\omega^2 = N_e/\omega^2,
\quad \omega \sim \infty ,
\end{equation}
indicating that a representation of $\alpha (i\omega)$ that only
includes the valence electron excitations $(N_e =1)$ will not have the
proper high-frequency tail, thereby leading to an inaccurate $C_3$
coefficient from the integral in eq.~(\ref{C3}).  We ensured that a
semi-empirical calculation of $C_3$ for Na~\cite{KhaBabDal97} included
the contribution of the (ionic) core electrons by requiring the
oscillator strength distribution to satisfy the Thomas-Reiche-Kuhn sum
rule.  A related argument concerning inclusion of all virtual
excitations was presented by Barton~\cite{Bar74}, who showed that a
two-level atomic model is inadequate to describe energy shifts of an
atom near a wall.

\section{\label{ideal}Ideal walls}

For an atom and a perfectly conducting wall,
an expression for the 
potential that is valid from small $R$ to asymptotically
large $R$ is available~\cite{CasPol48,DzyLifPit61,TikSpr93a},
\begin{equation}
\label{AtM}
V_{\mathrm{At}M} (R) = -\frac{C_3f_3 (R)}{R^3} ,
\end{equation}
where the dimensionless retardation coefficient is
\begin{equation}
\label{f3}
f_3 (R) = \frac{1}{8 C_3 \pi \alpha_{\mathrm{fs}} R} 
       \int_0^\infty dx\; e^{-x}\alpha_d (ix/2\alpha_{\mathrm{fs}} R) 
         [\case{1}{2}x^2+x+1]  ,
\end{equation}
and the subscript $\textrm{At}\textit{M}$ denotes the atom-metal wall
interaction~\cite{TikSpr93a}, with $\alpha_{\mathrm{fs}}$ the fine
structure constant.  Eq.~(\ref{AtM}) has the unretarded result
eq.~(\ref{small-R}) as its limit for small $R$ and the Casimir-Polder
result eq.~(\ref{CP}) as its limit for large $R$.

For a wall with a dielectric constant $\epsilon$, the potential can be
written~\cite{DzyLifPit61,TikSpr93a}
\begin{equation}
\label{AtD}
V_{{\rm At}D} (R,\epsilon) =
  -\frac{\alpha_\mathrm{fs}^3}{2\pi} \int_0^\infty d\xi \xi^3 \alpha_d (i\xi)
   \int_1^\infty dp \exp (-2\xi R p\alpha_\mathrm{fs}) H [p,\epsilon (i\xi)] ,
\end{equation}
where
\begin{equation}
H (p,\epsilon) = \frac{s-p}{s+p} 
        + (1-2p^2)\frac{s-\epsilon p}{s+\epsilon p} 
\end{equation}
and 
\begin{equation}
s = (\epsilon - 1 +p^2)^{1/2} 
\end{equation}
and 
the subscript $D$ denotes the dielectric wall.

Accurate Lennard-Jones coefficients and dynamic dipole
polarizabilities for He$(2\,^3\!S)$ have been used, for example, to
theoretically analyze matter wave interference in an atomic
trampoline~\cite{MarCogSav00}, for a comparison to the experimental
results of atomic diffraction from a silicon nitride
grating~\cite{BruFouGri02}, and in analysis of quantum reflection of
atoms off of a flat polished Si surface~\cite{ObeTasShi05}.

\begin{figure}[h]
\includegraphics[width=22pc]{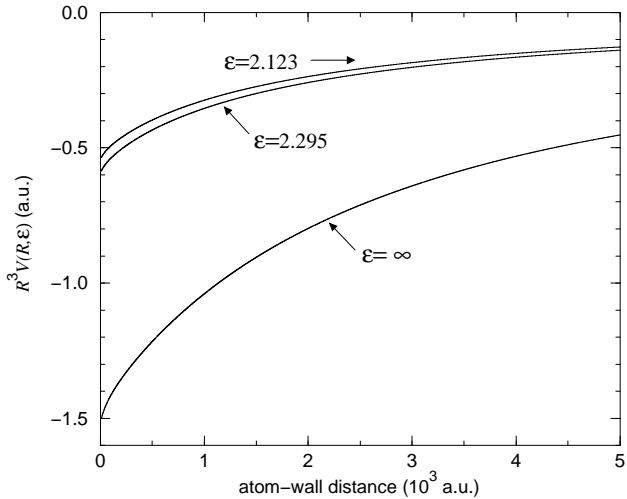}\hspace{2pc}%
\begin{minipage}[b]{14pc}\caption{\label{Liwall}Accurate
    values of $ V_{\mathrm{At}D} (R,\epsilon)$ for a Li atom and walls
    with 
    $\epsilon=2.123$, 
    $\epsilon=2.295$, and for a perfectly conducting wall
    $(\epsilon=\infty)$\protect\cite{YanDalBab97},
   in atomic units.}
\end{minipage}
\end{figure}
In fig.~\ref{Liwall} plots of accurate values of the atom-wall
potentials $V_{{\rm At}D} (R,\epsilon)$, eq.~(\ref{AtD}), for a Li
atom and a wall with $\epsilon=2.123$, a wall with $\epsilon=2.295$,
and for a perfectly conducting wall
$(\epsilon=\infty)$~\cite{YanDalBab97} are presented.
Eq.~(\ref{AtD}) can be readily evaluated for a wall characterized by a
dielectric constant once an accurate dynamic electric dipole
polarizability function is available.  For short range, the effect of
the dielectric wall on $C_3$ is a reduction by a factor,
\begin{equation}
 V_{{\rm At}D} (R,\epsilon) \rightarrow  -\frac{1}{4\pi R^3} 
 \int_0^\infty \,d\omega \alpha_d (i\omega)
 \frac{\epsilon (i\omega)-1}{\epsilon (i\omega)+1}  ,\qquad
 \mathrm{small}\;R.
\end{equation}
For a fixed dielectric constant, $C_3$ is reduced by the factor
$[(\epsilon-1)/(\epsilon+1)]$, as is evident in fig.~\ref{Liwall}.  A
similar expression describes the reduction of the Casimir-Polder
asymptotic potential for a dielectric wall~\cite{DzyLifPit61}.

\section{\label{beyond-ideal}Beyond ideal}

Thus far, the consideration of the atom has been limited to the
electric dipole polarizability and the surface has been considered to
be either a perfect conductor or a material with a fixed dielectric
constant.  Actual surfaces could consist of real metals or dielectrics
with frequency dependent properties, possess a nonzero surface
temperature, have geometries deviating from a plane, and consist of
layers of substrates. Also of interest is the consideration of higher
electric multipoles or magnetic interactions and the treatment of an
atom in an excited state or of a molecule.  We now address these
issues.

\subsection{Finite temperature and frequency-dependent dielectric constant}

Expressions for the atom-wall interaction, for a wall at finite
temperature, valid at all but very short distances are available,
cf.~\cite{Par74,MarToi82a,BoSer00,BabKliMos04,AntPitStr04,CarKliMos05}.
In considering finite temperature, a new distance scale appears, the
thermal de Broglie wavelength $\hbar c/k_B T$ of the photons.  More
formally, the energy in each photon field mode is replaced by the the
energy including the thermal photons
\begin{equation}
\textstyle{\frac{1}{2}}\hbar c k \rightarrow
\textstyle{\frac{1}{2}}\hbar c k + \hbar c k [\exp (\hbar c k/k_B T)-1]^{-1} .
\end{equation}
The expression for the atom-wall potential 
at finite temperature accordingly involves
a Matsubara summation over frequencies, cf.~\cite{NinParWei70}.
In the classical limit of high temperature, where the real photons dominate
the virtual photons,
the atom-wall potential becomes
\begin{equation}
\label{classical}
V(R,T) \sim  -\frac{1}{4} k_B T \alpha_d (0) /R^3 ,
\end{equation}
and note the absence of $\hbar$.  Spruch~\cite{Spr02} has shown that
the classical limit $RT\sim\infty$ simply arises from the replacement
of $\frac{1}{2}\hbar k c$ by $k_B T$. This replacement combined with arguments
similar to those leading to eq.~(\ref{physical}), where we replaced
$\textbf{E}_b^2(\omega)$ with $\hbar \omega/{\rm V}$ to account for
virtual photons, would lead to the essential properties of
eq.~(\ref{classical}), see also Ref.~\cite{Boy75}.
For a dielectric material eq.~(\ref{classical}) is diminished
by the factor $[(\epsilon-1)/(\epsilon+1)]$~\cite{AntPitStr04}.

For surface temperatures of 300~K, evaluations of the interaction
potentials for small distances are available for He$(2\,^3\!S)$ or Na
atoms and an Au, Si, or SiO${}_2$ wall~\cite{CarKliMos05} and
evaluations for large distances are available for He$(2\,^3\!S)$, Na,
or Cs atoms and an Au wall~\cite{BabKliMos04}, and Rb atoms and a
sapphire surface~\cite{AntPitStr04}.  In addition, there is study of
the interaction potential of H atoms near an Ag
surface~\cite{BoSer00}.

One ingredient of these calculations that I have not covered in this
tutorial is the frequency dependent dielectric constant $\epsilon
(i\omega)$---another topic with a vast literature.  The reader is
referred to Refs.~\cite{AntPitStr04} and \cite{CarKliMos05},
respectively, for example treatments of the $\epsilon (i\omega)$
function of sapphire and Au.

\begin{figure}[h]
\includegraphics[width=18pc]{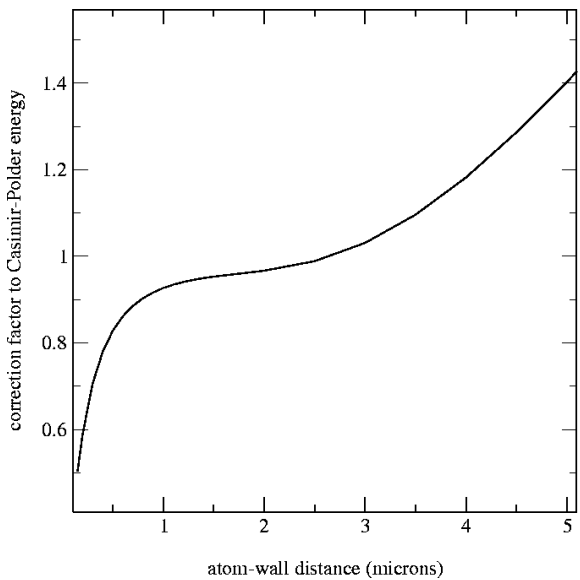}\hspace{6pc}%
\begin{minipage}[b]{14pc}\caption{\label{HeAuwall}Accurate
    values of the dimensionless correction factor
   to the Casimir-Polder energy, eq.~(\protect\ref{CP}), 
   for a  He$(2\,^3\!S)$ atom and 
   a gold wall at 300~K~\protect\cite{BabKliMos04}.}
\end{minipage}
\end{figure}

The evaluation of the atom-wall potential at finite temperature for a
real wall is illustrated in fig.~\ref{HeAuwall}, where
the dimensionless multiplicative correction factor to the
Casimir-Polder potential is plotted for a He$(2\,^3\!S)$ atom and a Au
wall at 300~K~\protect\cite{BabKliMos04}.  The factor is the ratio 
of the atom-wall potential to the asymptotic
Casimir-Polder potential of eq.~(\ref{CP}), $-
\textstyle{\frac{3}{8\pi}}\hbar c \alpha_d(0)R^{-4}$.  The linear
dependence of the correction factor at small $R$ indicates the
Lennard-Jones potential eq.~(\ref{small-R}) is a good approximation
and the linear dependence of the correction factor at large $R$
indicates that the classical potential tail eq.~(\ref{classical}) is a
good approximation.  The roughly flat behavior for separations between
1 and 3 microns indicates the applicability of the Casimir-Polder
potential, eq.~(\ref{CP}).

\subsection{Surface roughness and layers}

Surface roughness effects on the interaction of an atom and a wall
have been considered, cf.  Refs.~\cite{MarToi82a,MarToi82b,BezKliRom00}.
The interaction of an atom with a substrate consisting of mutiple layers 
was investigated in Ref.~\cite{ZhoSpr95}.
For thin layers, the power law describing the potential
is predicted to be non-integer for certain cases, cf.
Refs.~\cite{BarKya89} and \cite{BoSer00}.

\subsection{Higher multipoles}

In addition to the long-range potential arising from the induced
electric dipole moment eq.~(\ref{small-R}), there is an induced
quadrupole moment. The interaction potential will be weaker and is
expected to drop off as the inverse fifth power.  Some expressions and
evaluations of coefficients are available in the
literature, cf. Ref.~\cite{HutFowZar86}.

\subsection{Molecules}

Studies of the interactions of diatomic molecules with surfaces along
the lines of recent work with atoms is of interest, as advances in
ultra-cold molecule science are continuing~\cite{DoyFriKre04}.  There
will be two independent components for a diatomic molecule, similarly
to a $P$-state atom interacting with a surface~\cite{BloDuc05}.
Theoretical expressions and evaluations of molecule-surface
interaction coefficients treating the asymmetric part were given
in Refs.~\cite{ShiRasKus74,HarFei82,GirGirSil88,HulRydLun99}.

\section{Other aspects}

In the study of atom-wall interactions the surfaces are usually
empirically described.  Further developments might lead to \textit{ab
  initio} calculation of 
surface material properties and atomic properties
simultaneously, perhaps with density functional
theory~\cite{HulRydLun99} or path-integral methods~\cite{EmiBus04}
Another intriguing area of research is the repulsive Casimir force,
which occurs in the interaction between a fluctuating electric dipole
moment and a fluctuating magnetic moment~\cite{FeiSuc70}.

\ack 
The Institute for Theoretical Atomic, Molecular, and Optical
Physics is supported by a grant from the NSF to the Smithsonian
Institution and Harvard University.  

\section*{References}

\end{document}